\documentclass[runningheads]{svmult}

\usepackage{makeidx}   
\usepackage{graphicx}  
\usepackage{subeqnar}  
\usepackage{multicol}  
\usepackage{physprbb}  
\makeindex             

\def\xHFC{HFC $E_{\rm pk} (\Phi)$}
\def\HIC{HIC $E_{\rm pk} (N)$ }
\def\HFC{HFC $E_{\rm pk} (\Phi)$ }

\def\Epkmax{E_{\rm pk,0}}

\def\NE{N_{\rm E}}

\def\F0{F_{\rm 0}}
\def\t0{t_{\rm 0}}

\def\td{\tau_{\rm d}}

\def\P0{\Phi_{\rm 0}}
\def\tP0{\tilde{\Phi_{\rm 0}}}
\def\E00{E_{\rm pk,0}}

\def\N0{N_{\rm 0}}

\newcommand{\ltsima} {$\; \buildrel < \over \sim \;$}
\newcommand{\gtsima} {$\; \buildrel > \over \sim \;$}
\newcommand{\lta} {\lower.5ex\hbox{\ltsima}}
\newcommand{\gta} {\lower.5ex\hbox{\gtsima}}
\begin{document}
\title*{A Variety of Decays of Gamma-Ray Burst Pulses}
\toctitle{A Variety of Decays of Gamma-Ray Burst Pulses}

\titlerunning{A Variety of Decays of GRB Pulses}
%
\author{Felix Ryde\inst{1,}\inst{2}
\and Roland Svensson\inst{2}}
\authorrunning{Ryde \& Svensson}
%
%
\institute{Center for Space Science and Astrophysics, Stanford University, Stanford CA 34305
\and Stockholm Observatory, SE-133 36 Saltsj\"obaden, Sweden}

\maketitle              

\section*{Introduction}

The main target of this study is the GRB light curve during the decay phase 
of long, bright pulses. As shown by Ryde \& Svensson (2000; hereafter
RS00) approximately half of these decays can be 
described by a power law $\propto$ 1/(time). This happens for cases when 
the hardness-fluence correlation (HFC) is an exponential function, 
$E_{\rm pk} (\Phi) \propto e ^{-\Phi / \Phi_{\rm 0}}$, and the
hardness-intensity correlation (HIC) is a power law, 
$E_{\rm pk} (N) \propto (N/\N0)^{\delta}$. Here, $N(t)$ is the
instantaneous photon flux, $E_{\rm pk}(t)$
is the corresponding photon energy, at which the $E^2 
\NE$-spectrum peaks and is used as a measure of
the spectral hardness, and the photon fluence is defined by $\Phi (t) =
\int ^t N(t')\,dt'$. These most commonly assumed 
correlations were found by Liang \&
Kargatis (1996; HFC) and Golenetskii et al. (1983; HIC).

There obviously exists a large group of GRB pulses which decay in 
a different way. In this paper, we search for alternative 
descriptions of the spectral/temporal evolution.
We use the complete sample of long pulses in strong bursts
presented in Ryde \& Svensson (2001) consisting of 25 pulses within 
23 bursts observed by BATSE on the {\it CGRO}
during its entire mission (1991 -- 2000). The spectral 
analysis of the LAD/HERB data  ($\sim 25-1900$ keV) 
was performed with the WINGSPAN/MFIT package (Preece et~al.\ 1996). 
For each time bin the photon spectrum with the
background subtracted was determined using  
the Band et al. (1993) function with both its 
power law indices left free to vary.
The instantaneous, integrated photon flux, $N(t)$, was
found by integrating  the modeled photon spectrum
over the available energy band.

\section*{Other Types of Behaviors}

There is no consensus on what shape the pulse decays have. 
Both power law and stretched exponential decays have been used.
Guided by the findings of RS00 we study the following
generalized power law decay:
\begin{equation}
N(t)= \frac{\N0}{(1+t/\tau)^n},\label{recnn}
\end{equation}
where $t$ is taken from the start of the decay, when  $[N(t),E_{\rm
pk}(t)]=[\N0,E_{\rm pk,0}]$ and the time constant 
 $\tau \equiv \delta \P0/\N0$, where $\P0$ is the
exponential decay constant of the exponential HFC and $\delta$ is the 
index of the power law HIC. The photon fluence associated with 
equation (\ref{recnn}) when $n$ differs from $1$ becomes
\begin{equation}
\Phi(t) = \frac{\N0 \tau}{n-1} \left\{ 1 - (1 + t /\tau ) ^{-(n-1)}
\right\}, \quad n \neq 1, \label{fluen}
\end{equation} 
which for $n$ larger than 1, converges 
to the asymptotic value $ f_{\rm 0} \equiv \N0 \tau  /(n-1)$.

Now, we consider two different alternatives. First, for GRB pulse 
light curves whose decays follow equation (\ref{recnn}), 
and for which the \HFC is an exponential, the \HIC will
follow  
\begin{equation}
E_{\rm pk} (N) = \Epkmax {\rm exp}\left\{\frac{ f_{\rm 0}}{\P0} 
\left[ \left(\frac{N}{\N0}\right) ^{(n-1)/n} -1 \right]
\right\} , \quad   n \neq 1. \label{HICn}
\end{equation}
When $\ln (\N0/N) << 2 n/\vert n-1\vert$ the HIC
approaches a power law
with the exponent $\delta/n$, which becomes identical to
the original power law HIC, when $n$ tends to
$1$. 
On the other hand, if the \HIC actually is a power law
then the \HFC will follow
\begin{equation}
E_{\rm pk} (\Phi) = \E00 \left(  1-\frac{\Phi}{f_{\rm 0}} \right)^{n \delta
/(n-1)} ,\qquad n \neq 1,\label{HFCn}
\end{equation}
which behaves similarily to the exponential HFC as $n$ tends to 1.

We also fitted the decays with a  stretched exponential:  
$N 
\propto \exp(-(t/\td)^{\nu})$, 
where 
$\td$ is the time constant for the decay
phase and $\nu$ is the peakedness parameter. This function is 
the most commonly assumed pulse shape used so far (e.g. Norris et~al. 1996).

Our study showed that, first, a power law gives a better description of the
pulse decays than a stretched exponential and, second, the
power law index  (Eq. 1) has a bimodal distribution in that there are two
preferred values $n=1$ and $n \sim 3$ (See Fig.1).
The sample is divided into approximately two equally large sets by $n \sim 2$.
For the 11 pulse decays with $n$ larger than $2$, we found 
that for each case either the \HIC or the \HFC is still valid,
while the other corresponding correlation is different, and thus
described by a new function. To be able to get constrained fits on all
cases we had to freeze the values of $\N0$ and $n$ to the values
obtained from the fits of the light curve.

\begin{figure}[b]
\begin{center}
\includegraphics[width=50mm]{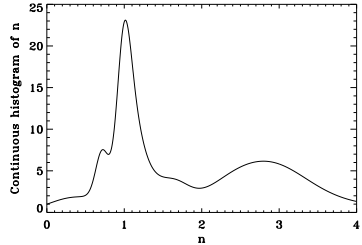}
\end{center}
\caption[]{Continuous histogram of the power law 
index $n$.}
\label{fig3}
\end{figure}

Six out of these eleven cases are, however, good enough for $n$ to be 
constrained. Four out of these gave $n$-values that were the same to 
within the errors as the values obtained from fitting the light curve. 
In the last two cases the errors in the $n$-values were 
so large that no certain conclusion could be drawn. 
In all of these four cases the power law \HIC is valid.

This suggests that the important relations for a GRB 
pulse decay are the power law \HIC and the light curve, $N(t)$.
The power law correlation between the hardness and the intensity
is valid independent of the shape of the light curve. The \xHFC,
on the other hand, is different for different light curve behaviors
according to equation (\ref{HFCn}), since the fluence is the time integral of 
the instantaneous flux.

\begin{figure}[b]
\begin{center}
\includegraphics[width=110mm]{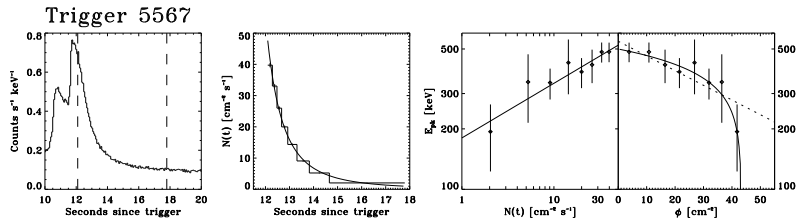}
\end{center}
\caption[]{Spectral and temporal behaviour of GRB960807 (BATSE trigger 5567).}
\label{fig2}
\end{figure}

In Figure 2 one of these cases, GRB960807, is presented.
The first panel shows the DISCSC data
(all four energy channels) and indicates the time interval 
studied and the second panel shows the light curve with the LAD
HERB data in the chosen time binning. The best fit is indicated with a solid
curve. The two left-hand panels, show the
correlations, the \HIC in panel 3 and the \HFC in panel 4. The fit of an
exponential HFC is shown by a dashed line. 

\section*{ACKNOWLEDGMENTS}

We are grateful to the GROSSC at NASA/GSFC 
for providing the HEASARC Online Service.
We acknowledge financial support from the Swedish Foundation for 
International Cooperation in Research and Higher Education (STINT) 
and the CF Liljevalch J:or Fund at Stockholm University.

\end{document}